\documentclass[twocolumn,showpacs,amsmath,amssymb,prl,nofootinbib]{revtex4}
\usepackage{graphicx}
\usepackage{dcolumn}
\usepackage{bm}
\usepackage{psfig}

\newcommand{\wf}{$\omega\phi $}
\newcommand{\psirwf}{$J/\psi \to \gamma \omega\phi  $}
\newcommand{\wppp}{$\omega\to\pi^+\pi^-\pi^0$}
\newcommand{\fkk}{$\phi\to K^+K^-$}

\begin{document}

\title{\bf \boldmath Observation of a near-threshold enhancement in the
\wf~mass spectrum from the doubly OZI suppressed decay \psirwf}

\author{
M.~Ablikim$^{1}$,              J.~Z.~Bai$^{1}$,               Y.~Ban$^{12}$,
J.~G.~Bian$^{1}$,              X.~Cai$^{1}$,                  H.~F.~Chen$^{17}$,
H.~S.~Chen$^{1}$,              H.~X.~Chen$^{1}$,              J.~C.~Chen$^{1}$,
Jin~Chen$^{1}$
% ...
\\
\vspace{0.2cm}
(BES Collaboration)\\
\vspace{0.2cm}
{\it
$^{1}$ Institute of High Energy Physics, Beijing 100049, People's Republic of China\\
% ...
$^{12}$ Peking University, Beijing 100871, People's Republic of China\\
% ...
$^{17}$ University of Science and Technology of China, Hefei 230026, People's Republic of China\\
% ...
}}

\vspace{0.4cm}

\date{\today}

\begin{abstract}
  {An enhancement near threshold is observed in the $\omega \phi$
    invariant mass spectrum from the doubly OZI suppressed decays of 
    $J/\psi \to \gamma \omega \phi$, based on a sample of $5.8 \times 10^7$
    $J/\psi$ events collected with the BESII detector.  A partial wave
    analysis shows that this enhancement favors $J^P = 0^+$, and its
    mass and width are $M = 1812^{+19}_{-26} ~{\rm (stat)} \pm 18$ (syst)
    MeV/$c^2$ and $\Gamma = 105\pm20 ~{\rm (stat)} \pm 28$ (syst) MeV/$c^2$.
    The product branching fraction is determined to be $B(J/\psi\to
    \gamma X)\cdot B(X\to\omega\phi) = (2.61\pm0.27 ~{\rm (stat)} \pm
    0.65$ (syst)) $\times 10^{-4}$.}
\end{abstract}

\pacs{12.39.Mk, 13.20.Gd, 13.30.Ce, 14.40.Cs}

\maketitle


QCD predicts a rich spectrum of $gg$ glueballs, $qqg$ hybrids and
$qq\bar{q}\bar{q}$ four quark states along with the ordinary
$q\bar{q}$ mesons in the 1.0 to 2.5 GeV/$c^2$ mass region. Radiative 
$J/\psi$ decays provide an excellent laboratory to search for these 
states. Until now, no clear experimental signatures for glueballs or 
hybrids have been found.

Recently, anomalous enhancements near threshold in the invariant
mass spectra of $p \bar p$ and $p \bar \Lambda$ pairs were observed 
in $J/\psi \to \gamma p \bar p$ \cite{gppb} and $J/\psi \to p K \bar
\Lambda$ \cite{pkl} decays, respectively, by the BESII experiment.
These surprising experimental observations stimulated many theoretical
speculations. Therefore it is of special interests to search for
possible resonances in other baryon-antibaryon, baryon-meson, and
meson-meson final states.

Systems of two vector particles have been intensively studied for
signatures of gluonic bound states. Pseudoscalar enhancements in
$\rho\rho$ and $\omega\omega$ final states have been seen in radiative
$J/\psi$ decays~\cite{mark1,dm1,mark2,dm2}, and resonant $\phi\phi$
structures have also been observed near threshold in $\pi p$
scattering experiments~\cite{ff}. The radiative $J/\psi$ decay \psirwf~
is a doubly OZI suppressed process, and its production ratio should be
suppressed by at least one order of magnitude.  Therefore, the
measurement of this decay and the search for possible resonant
states will provide useful information on two vector meson systems.
MARKIII collaboration \cite{mark3phi} studied $J/\psi \to \gamma \omega \phi$
decays, but did not find clear structures in the 
$\omega\phi$ invariant mass spectrum. The final states of $\omega\phi$ 
were also observed in photon-photon collisions by 
ARGUS \cite{argus1,argus2} experiment and the cross sections were 
measured \cite{argus2}. 

In this letter, we report on the measurement of the doubly OZI
suppressed $J/\psi \to \gamma \omega \phi$ decay and an enhancement 
near threshold in the $\omega \phi$ invariant mass spectrum, using  
$5.8 \times 10^7$ $J/\psi$ events collected with the  upgraded Beijing
Spectrometer (BESII) at the Beijing Electron-Positron Collider (BEPC). 
BESII is a large solid-angle magnetic spectrometer  that is described 
in detail in Ref. \cite{BESII}. 


The \psirwf ~(\wppp, \fkk) candidate events are required to have four 
charged tracks, each of which is well fitted to a helix that is within 
the polar angle region $|\cos\theta|<0.8$ in the main drift chamber (MDC) 
and has a transverse momentum larger than 50 MeV/$c$. The total charge 
of the four tracks is required to be zero. For each track, the time-of-flight 
(TOF) and specific ionization ($dE/dx$) measurements in MDC are combined 
to form particle identification Chi-squares for the $\pi$, K, and $p$ 
hypotheses, and the overall Chi-square is determined by adding those of 
the individual tracks.  The $K^+K^-\pi^+\pi^-$ combination is chosen as 
the combination with the smallest combined particle identification 
Chi-square, $\chi^2(\pi^+ \pi^- K^+ K^-)$, which is required to be smaller 
than $\chi^2(\pi^+ \pi^- \pi^+ \pi^-)$ (for the $\pi^+\pi^-\pi^+\pi^-$ 
hypothesis) and  $\chi^2(K^+ K^- K^+ K^-)$ (for the $K^+K^-K^+K^-$ hypothesis)
to remove the background with $\pi^+\pi^-\pi^+\pi^-$ and $K^+K^-K^+K^-$ 
final states.

Candidate photons are required to have an energy deposit in the barrel shower 
counter (BSC) greater than 40 MeV, to be isolated from charged tracks by more 
than $10^{\circ}$,  and to have the difference of angle between the cluster 
development direction in the BSC and the photon emission direction less than 
$60^{\circ}$. The number of photons is required to be in the range from 3 to 6.

A five-constraint (5C) energy-momentum conservation kinematic fit is made 
under the $J/\psi \to \gamma K^+K^-\pi^+\pi^-\pi^0$ hypothesis with the 
invariant mass of the $\gamma \gamma$ pair associated with the $\pi^0$ being 
constrained to $m_{\pi^0}$. The combination of gammas with the largest 
probability is chosen as the best combination, and events with probability 
larger than $1\%$ are retained.


To remove backgrounds from $J/\psi \to K^+K^-\pi^+\pi^-\pi^0$ and $J/\psi 
\to K^+K^-\pi^+\pi^-\pi^0\pi^0$, a 5C kinematic fit to the $J/\psi \to 
K^+K^-\pi^+\pi^-\pi^0$ hypothesis and a 6C kinematic fit to $J/\psi \to 
K^+K^-\pi^+\pi^-\pi^0\pi^0$ (if the number of good photons is greater than 4) 
are performed, and the probabilities are required to be less than that from 
the 5C fit to the signal channel. To remove background where the $\pi^0$ is 
falsely reconstructed from a high energy photon and a second spurious shower, 
the requirement $|E_{\gamma 1}-E_{\gamma 2}|/ |E_{\gamma 1}+E_{\gamma 2}|<0.90$ 
is applied to the photons forming the $\pi^0$. Here, $E_{\gamma 1}$ and
$E_{\gamma 2}$ are the energies of the two photons.
\begin{figure}[htbp]
\centering
{
\includegraphics[width=4.5cm,height=3.5cm]{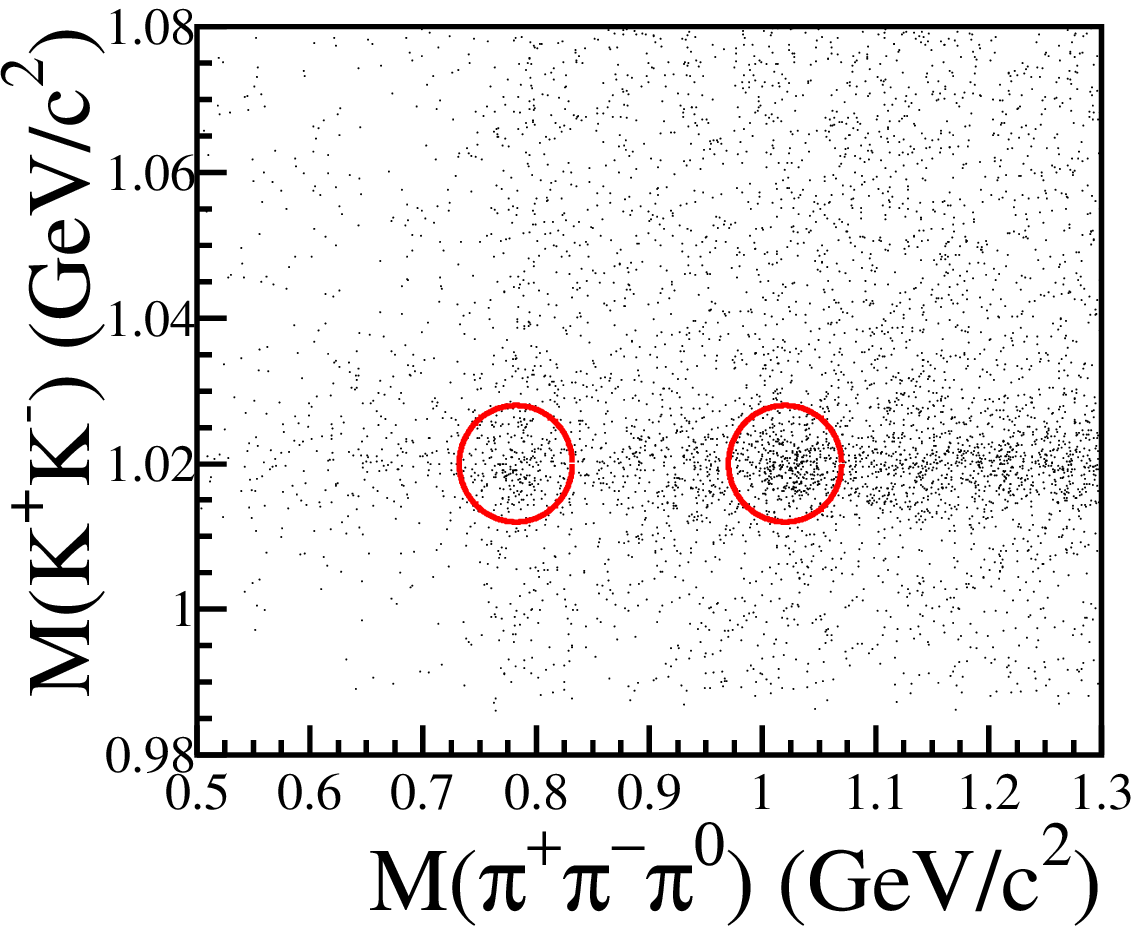}
\put(-100,80){(a)}}
\centering
{\includegraphics[width=4.2cm,height=3.5cm]{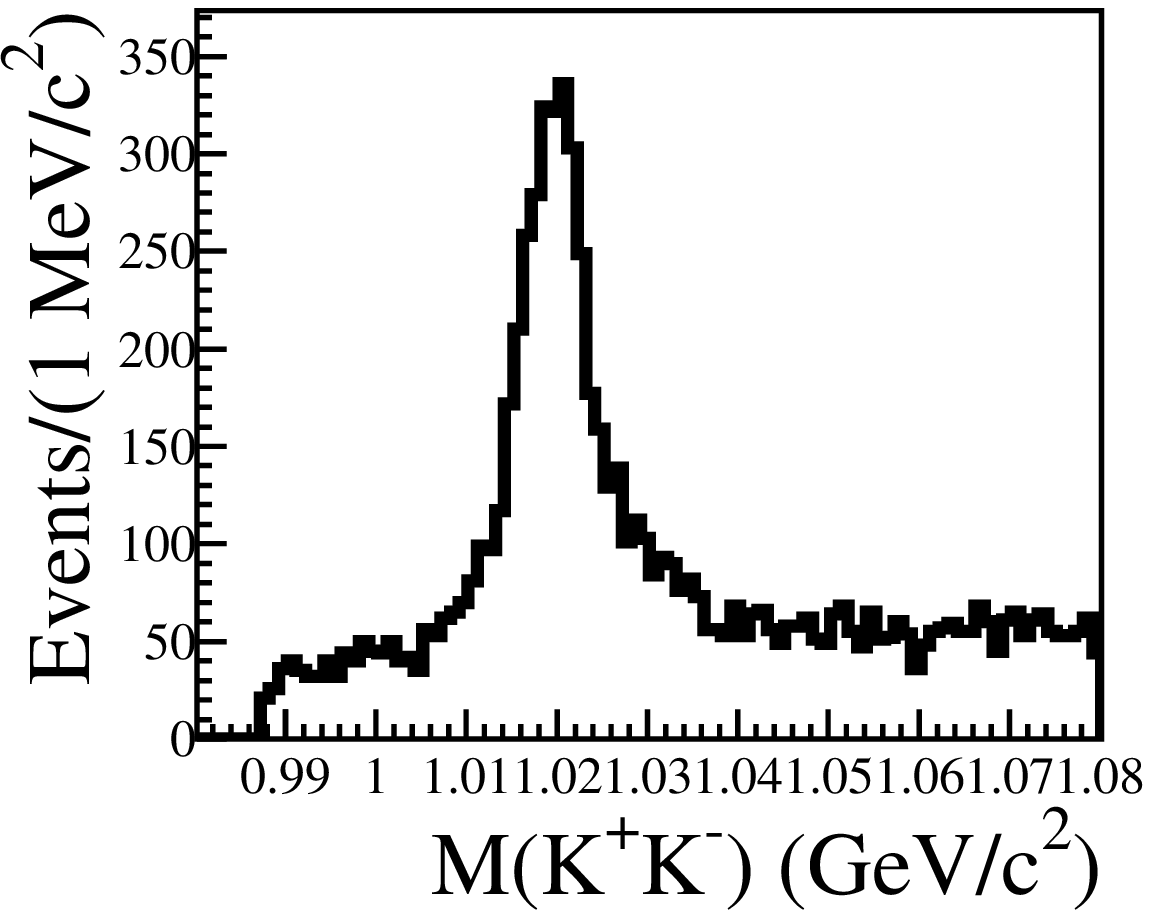}
\put(-25,80){(b)}
\includegraphics[width=4.2cm,height=3.5cm]{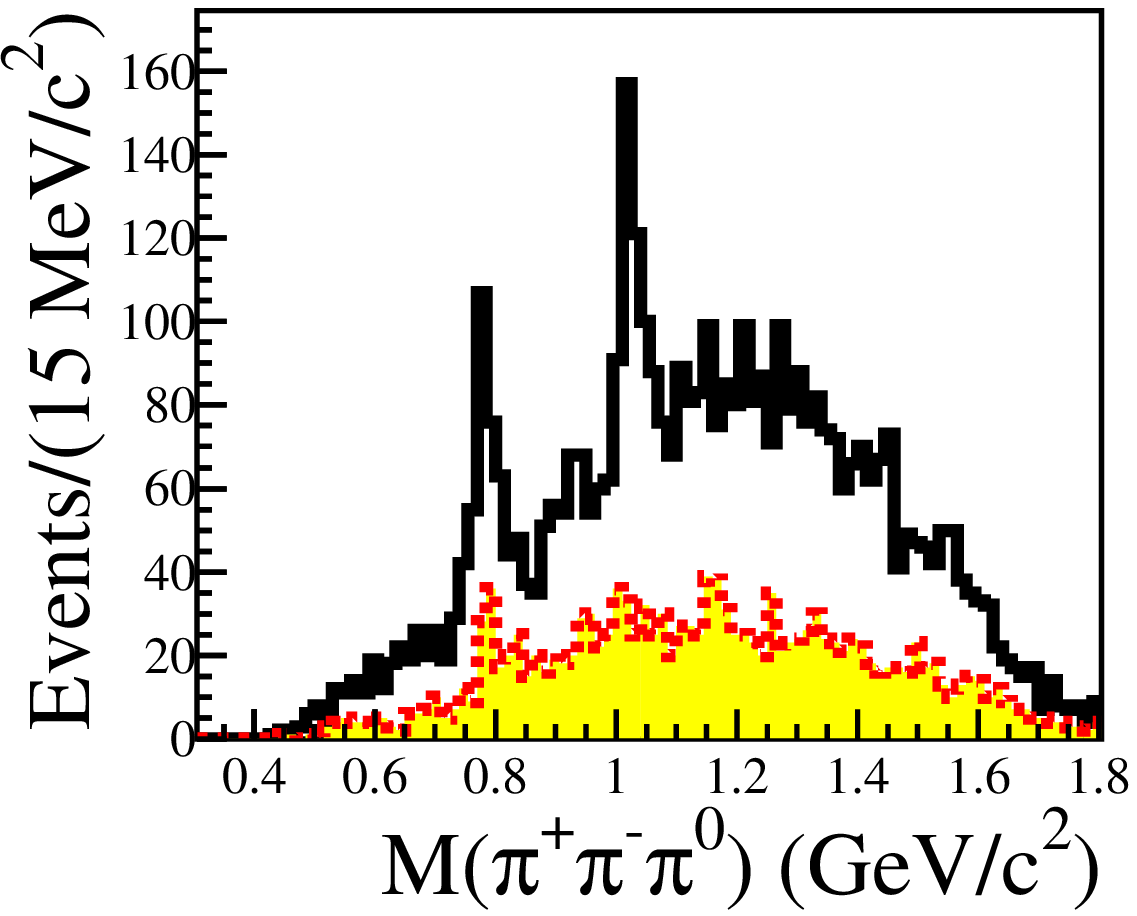}
\put(-25,80){(c)}}
{\includegraphics[width=4.2cm,height=3.8cm]{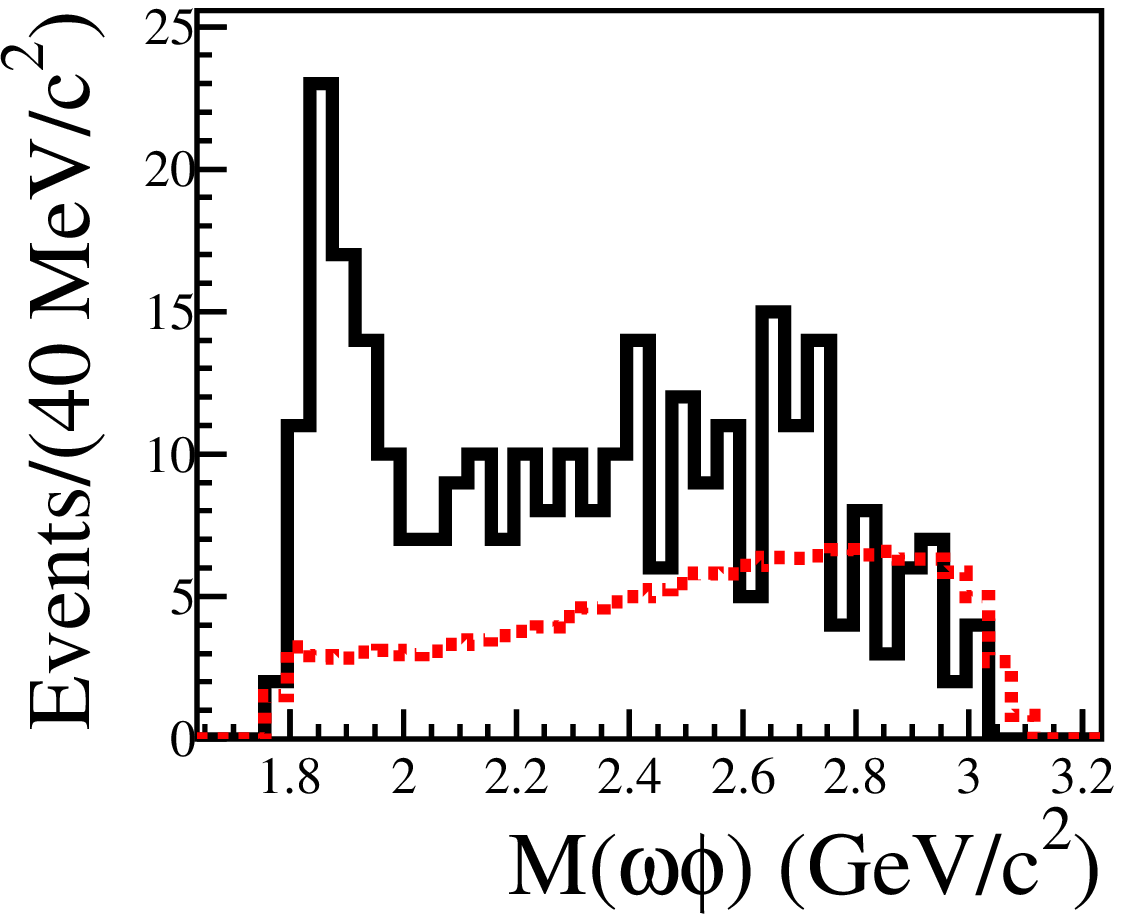}
\put(-25,85){(d)}
\includegraphics[width=4.2cm,height=3.8cm]{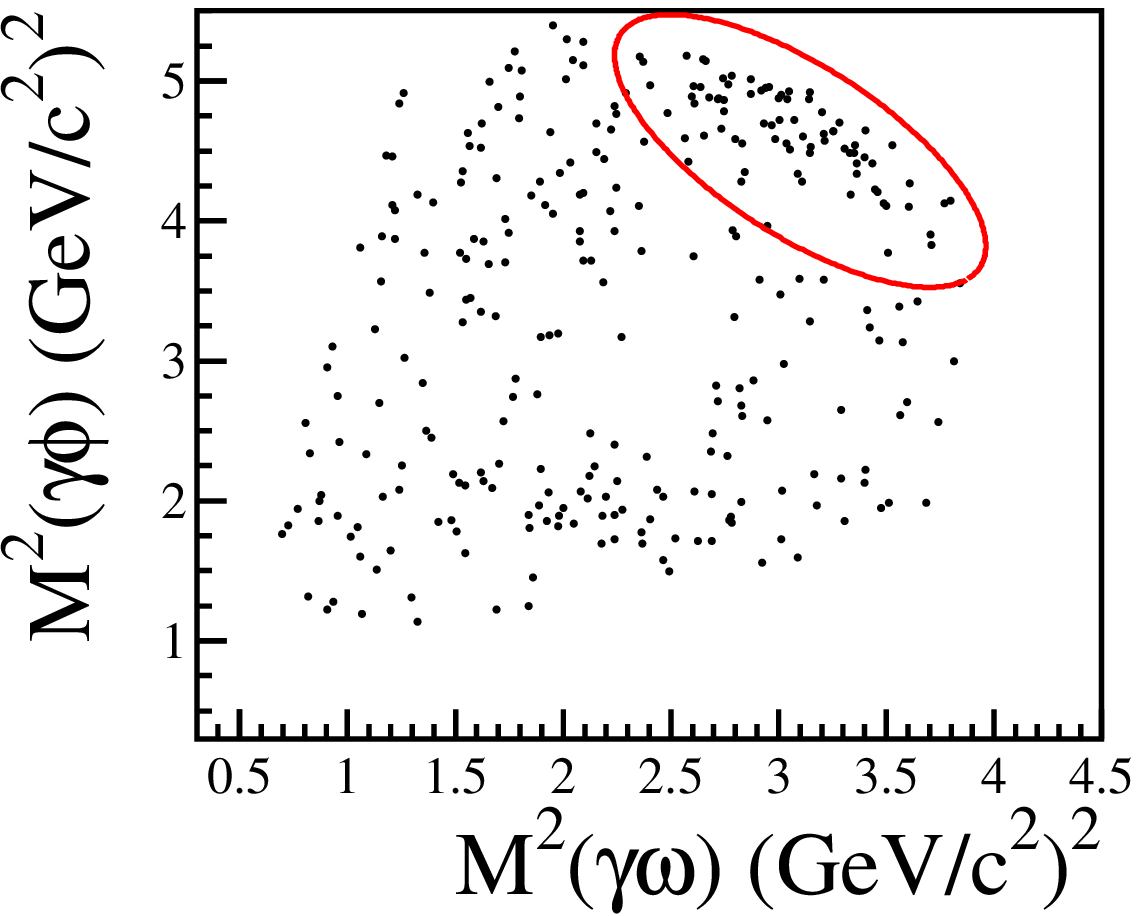}
\put(-25,85){(e)}}
\caption{
(a) The scatter plot of the  $m_{K^+K^-}$ versus the $\pi^+\pi^-\pi^0$ 
invariant mass.
(b) The $K^+K^-$ invariant mass distribution.
(c) The $\pi^+\pi^-\pi^0$ invariant mass distribution; the open histogram 
is for candidate events with $m_{K^+K^-}$ being in the $\phi$ range,
and the shaded histogram is for events with $m_{K^+K^-}$ being in the
$\phi$ sideband region.
(d) The $K^+K^-\pi^+\pi^-\pi^0$ invariant mass distribution for the 
$J/\psi\to \gamma \omega\phi$ candidate events. The dashed curve indicates 
the acceptance varying with the $\omega\phi$ invariant mass.
(e) Dalitz plot.}
\label{kk}
\end{figure}

Fig. \ref{kk}(a) shows the scatter plot of the $m_{K^+K^-}$ versus the 
$m_{\pi^+\pi^-\pi^0}$ invariant mass after applying the above selection 
criteria. Two clusters are clearly seen, which indicate the direct 
observation of the decays of $J/\psi \to \gamma \omega \phi$ and $\gamma 
\phi \phi$. The $K^+K^-$ invariant mass distribution is shown in 
Fig.~\ref{kk}(b), where the $\phi$ signal can be seen clearly. 
The $\pi^+\pi^-\pi^0$ 
invariant mass distribution of candidate events with $m_{K^+K^-}$ in the 
$\phi$ range ($|m_{K^+K^-}-m_{\phi}|<15$ MeV/$c^2$) is shown as the open
histogram in Fig.~\ref{kk}(c), where $\omega$ and $\phi$ signals are seen. 
The shaded histogram in Fig.~\ref{kk}(c) is the $\pi^+\pi^-\pi^0$ invariant 
mass spectrum recoiling against the $\phi$ sideband region 
(15 MeV/$c^2<|m_{K^+K^-}-m_{\phi}|<30$ MeV/$c^2$), where, only a very small 
$\omega$ signal is observed which comes from backgrounds such as 
$J/\psi\to\omega K^+K^-$, $\omega K^*K$, etc. Since the decays of $J/\psi
\to \omega\phi$ and $\pi^0 \omega \phi$ are forbidden by C-invariance, the 
294 observed $\omega\phi$ events present direct evidence for the radiative 
$J/\psi \to \gamma \omega\phi$ decay.

The histogram in Fig.~\ref{kk}(d) shows the $K^+K^-\pi^+\pi^-\pi^0$ invariant
mass distribution for events with $K^+K^-$ invariant mass within the nominal 
$\phi$ mass range ($|m_{K^+K^-}-m_{\phi}|<15$ MeV/$c^2$) and the  
$\pi^+\pi^-\pi^0$ mass within the $\omega$ mass range ($|m_{\pi^+\pi^-\pi^0}
-m_{\omega}|<30$ MeV/$c^2$), and a structure peaked near  $\omega\phi$ 
threshold is observed. There is no evidence of an $\eta_c$ signal in the 
$\omega \phi$ invariant mass spectrum. The dashed curve in the figure indicates
how the acceptance varies with invariant mass. The acceptance decreases as 
the invariant mass of $\omega \phi$ becomes smaller due to the decay of the 
kaon. The peak is also evident as a diagonal band along the upper right-hand 
edge of the Dalitz plot in Fig.~\ref{kk}(e). There is also a horizontal band 
near  $m^2_{\gamma K^+ K^-} = 2$ (GeV/$c^2$)$^2$ in the Dalitz  plot, which 
mainly comes from background due to $J/\psi \to \omega K^*K$. 

\begin{figure}[htbp]
\includegraphics[width=7.cm,height=6.0cm]{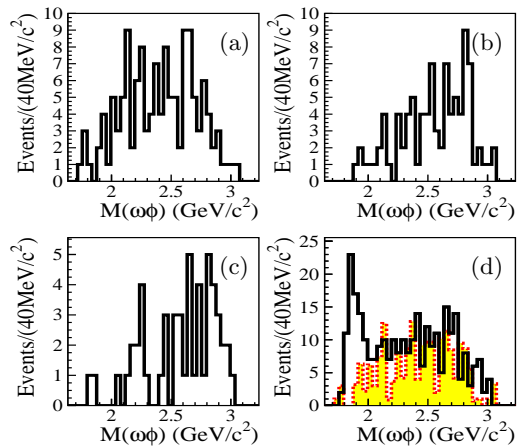}
\put(-120,150){(a)}
\put(-25,150){(b)}
\put(-120,65){(c)}
\put(-25,65){(d)}
\caption{The  $K^+K^-\pi^+\pi^-\pi^0$ invariant mass distribution for
(a) the events in the $\omega$ sideband;
(b) the events in the $\phi$ sideband;
(c) the events in the corner region;
(d) for events in the $\omega\phi$ range, as described in the text.
The shaded histogram in (d) represents the background distribution
obtained from the sideband evaluation.}
\label{sideband}
\end{figure}

To ensure that the structure at the $\omega\phi$ mass threshold is not due 
to background, we have studied potential background sources using both data 
and Monte Carlo (MC) data. Non-$\omega$ and non-$\phi$ background are studied 
using $\omega$ and $\phi$ sideband events. Fig. \ref{sideband}(a) shows the
$K^+K^-\pi^+\pi^-\pi^0$ invariant mass of events within the $\omega$ sideband 
(50 MeV/$c^2<|m_{\pi^+\pi^-\pi^0}-m{_\omega}|< 80$ MeV/$c^2$,  $|m_{K^+K^-}-
m_{\phi}|<15$ MeV/$c^2$), Fig. \ref{sideband}(b) shows the corresponding 
spectrum of events within the $\phi$ sideband ($|m_{\pi^+\pi^-\pi^0}-m{_\omega}|
<30$ MeV/$c^2$, 15 MeV/$c^2<|m_{K^+K^-}-m_{\phi}|<30$ MeV/$c^2$), and Fig. 
\ref{sideband}(c) shows the events in the corner region, which is defined as
(50 MeV/$c^2<|m_{\pi^+\pi^-\pi^0}-m{_\omega}|<80$ MeV/$c^2$, 15 MeV/$c^2<
|m_{K^+K^-}-m_{\phi}|<30$ MeV/$c^2$). The background is estimated by summing 
up the normalized backgrounds in Fig. \ref{sideband}(a) and Fig. \ref{sideband}(b) 
and subtracting that in Fig. \ref{sideband}(c), and it is shown as the shaded 
histogram in Fig. \ref{sideband}(d). No evidence of an enhancement near
$\omega\phi$ threshold is observed from the non-$\omega$ and non-$\phi$ 
background events.

Exclusive MC samples of $J/\psi$ decays which have similar final states are 
generated to check whether a peak 
near $\omega \phi$ mass threshold can be produced.
The main backgrounds come from $J/\psi \to \omega K^* K, K^*
\to K \pi^0$. About 45+/-17 $J/\psi \to \omega K^* K, K^*
\to K \pi^0$ events remain in the $\omega \phi$ invariant 
mass. However, they peak at the high mass region and do not produce a peak 
near the threshold. We also 
checked possible backgrounds with a 60 million Monte-Carlo $J/\psi \to 
anything$ sample, generated by the LUND-Charm model~\cite{chenjc}. None of 
the MC channels produces a peak near threshold in the $\omega\phi$ invariant 
mass spectrum. In addition,  the data taken at the $e^+e^-$ center of mass 
energy of 3.07 GeV, with a luminosity of $2272.8 \pm 36.4$ nb$^{-1}$ 
are used to check the continuum contribution. No events are survived.
As a check, the measurement of the branching fraction of
$J/\psi\to \gamma \phi\phi \to \gamma \pi^+\pi^-\pi^0 K^+K^-$ is performed, 
and the result is consistent with that from $J/\psi\to \gamma \phi\phi \to 
\gamma K^+K^-K^+K^-$, but with a larger error.


A partial wave analysis is used to study the spin-parity of the enhancement, 
denoted as $X$. The amplitudes are constructed with the covariant helicity 
coupling amplitude method~\cite{wuning}, and the maximum likelihood method 
is utilized. The decay process is described with sequential 2-body or 3-body 
decays: $J/\psi \to \gamma X$, $X \to\omega\phi$, $\omega\to\pi^+\pi^-\pi^0$, 
and $\phi\to K^+K^-$.  The resonance $X$ is parameterized by a Breit-Wigner 
with constant width, and the background is approximated by non-interfering 
phase space. The $\omega$ decay amplitude is not considered in the fit. The 
details of the likelihood function construction can be seen in Ref.~\cite{guozj}.

\begin{figure}[htbp]
\centering {\includegraphics[width=7.5cm,height=7.0cm]{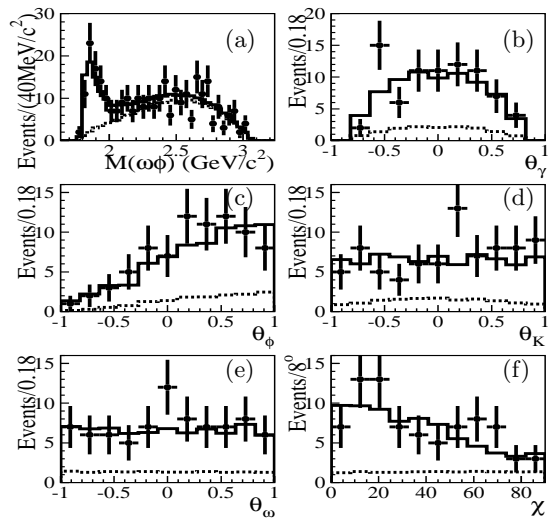}
\put(-25,180){(b)} \put(-130,180){(a)} \put(-25,120){(d)}
\put(-130,120){(c)} \put(-25,55){(f)} \put(-130,55){(e)} }
\vspace {-0.3cm}
\caption{Comparison between data and MC projections using the fitted parameters 
for the $\omega\phi$ invariant mass distribution and the angular distributions 
for the events with $\omega\phi$ invariant mass less than 2.0 GeV/$c^2$. Points
with error bars are data, the solid histogram is the MC projection, and the 
dashed line is the background contribution. 
(a) The $\omega\phi$ invariant mass distribution; 
(b) The polar angle of radiative photon ($\theta_{\gamma}$); 
(c) The polar angle of the $\phi$ in the $\omega\phi$ rest system ($\theta_{\phi}$). 
(d) The polar angle of kaon in the $\phi$ rest system ($\theta_K$) 
(e) The polar angle of the normal to the $\omega$ decay plane in the $\omega$
system. ($\theta_{\omega}$) 
(f) The $\chi$ distribution - the angle between azimuthal angles of the 
normal to the $\omega$ decay plane and the momentum of a kaon from $\phi$ 
decay in the $X$ rest system.}
\label{proj}
\end{figure}

When $J/\psi \to \gamma X$, $X \to \omega \phi$ is fitted with both the 
$\omega\phi$ and $\gamma X$ systems being $\mathcal{S}$-wave, which corresponds
to a $X = 0^{++}$ scalar state considering the C-parity of the $\omega\phi$ 
system, the fit gives the best log likelihood value of -79.55. The fit gives 
$95 \pm 10$ events with mass $M = 1812^{+19}_{-26}$ MeV/$c^2$, width $\Gamma = 
105 \pm 20$ MeV/$c^2$, and a statistical significance larger than 10 $\sigma$.
Fig.~\ref{proj}(a) shows the comparison of the $\omega\phi$ invariant mass 
distributions between data and MC projection with the fitted parameters. The 
comparisons of the angular distributions between data and MC projection for 
the events with the invariant mass less than 2.0 GeV/$c^2$ are shown in 
Fig.~\ref{proj}(b) - (f).

If the decay of the $\gamma X$ system in $J/\psi \to \gamma X$, $X \to \omega 
\phi$ is treated as $\mathcal{D}$-wave or a combination of both $\mathcal{D}$ 
and $\mathcal{S}$-wave, the mass and width of the $X$, as well as the log 
likelihood value do not change much.  However, if the decay of the $X$ to 
$\omega\phi$ is fitted with a $\mathcal{D}$-wave, the log likelihood value 
gets worse by about 40. This means that the orbital angular momentum of the 
$X \to\omega\phi$ decay can be well separated between $\mathcal{S}$-wave and
$\mathcal{D}$-wave, while it is difficult to determine in the $\gamma X$ 
system. A fit with $\mathcal{P}$-wave decays in both the $\omega\phi$ system 
and $\gamma X$ systems, corresponding to $X$ being a $0^{-+}$ pseudoscalar 
state, makes the log likelihood value worse by 58. Theoretically, the $0^{-+}$
hypothesis can be well separated from the $0^{++}$ hypothesis by the 
distributions of the polar angle of the $K^+$ ($\theta_{K^+}$) in the $\phi$ 
rest system, the polar angle of the normal to the $\omega$ decay plan 
($\theta_{\omega}$) in the $\omega$ rest system, and $\chi$, the angle between 
the azimuthal angles of the normal to the $\omega$ decay plane and the momentum
of a $K$ from $\phi$ decay in the $X$ rest system. We also tried to fit the
resonance $X$ with $2^{++}$ and $2^{-+}$ spin-parity hypotheses with all 
possible combinations of orbital angular momenta in the $\omega\phi$ and 
$\gamma X$ systems. The log likelihood values of the best fits are 
-Log($\mathcal{L}$) = -74.80 and -Log($\mathcal{L}$) = -63.81 for $2^{++}$ 
and $2^{-+}$ assignments, respectively. Although the fit is not too much 
worse for the $2^{++}$ case, there is only one free parameter for $0^{++}$, 
while there are four free parameters for $2^{++}$. Also for $2^{++}$, 
$\mathcal{D}$-wave is required in $X \to \omega\phi$; if only 
$\mathcal{S}$-wave is used in the fit, -Log($\mathcal{L}$) = -67, which is 
much worse than the final fit. Therefore we conclude that the $J^{PC}$ of the 
enhancement $X$ favors $0^{++}$.

Using the selection efficiency of 1.44\%, determined from Monte-Carlo
simulation, we obtain the product of the branching fractions as:
$$\mathcal{B}(J/\psi\to\gamma X)\cdot\mathcal{B}(X\to\omega\phi)
 = (2.61\pm0.27)\times10^{-4}.$$

Since phase space $J/\psi\to \gamma \omega\phi$ decays exist, fitting with 
an interfering phase space ($0^+$) is also performed, and the differences 
between fitting with non-interfering phase space for the mass, width, and 
branching ratio are $0.5\%$, $21.9\%$ and $13.8\%$, respectively. The 
differences will be included as systematic errors.

The systematic uncertainties on the mass and width come from the uncertainties 
in the background, the mass calibration, and the interference with phase space, 
as well as possible biases due to the fitting procedure. The latter are 
estimated from differences between the input and output masses and widths from 
MC samples, which are generated as $J/\psi \to \gamma 0^{++}, 0^{++} \to 
\omega\phi$ with $\mathcal{S}$-wave in both the $\omega\phi$ and $\gamma X$ 
systems. The uncertainties in the background include the uncertainty in the
amount of background as well as the treatment of the background in the fitting.  
We also tried to subtract the background determined from the sidebands in the 
fit instead of using the non-interfering or interfering phase space background, 
and the differences are taken as systematic errors.  The total systematic 
errors on the mass and width are determined to be 18 MeV/$c^2$ and 28 MeV/$c^2$,
respectively. The systematic errors in the branching fraction measurement 
mainly come from the efficiency differences between the Monte-Carlo simulation 
and data, which include the systematic uncertainties of the tracking efficiency, 
the photon detection efficiency, the particle identification efficiency, the 
kinematic fit, and the $\omega$ and $\phi$ decay branching fractions, the 
amount of background, MC statistics, the fitting procedures, different 
treatment of background, and the total number of $J/\psi$ events. The total
relative systematic error on the product branching fraction is $25 \%$.


In summary, the doubly OZI suppressed decay of $J/\psi\to\gamma \omega\phi, 
\omega\to\pi^+\pi^-\pi^0, \phi\to K^+K^-$ is studied. An enhancement near 
$\omega\phi$ threshold is observed with a statistical significance of more 
than 10$\sigma$.  From a partial wave analysis with covariant helicity 
coupling amplitudes, the spin-parity of the $X = 0^{++}$ with an 
$\mathcal{S}$-wave $\omega\phi$ system is favored. The mass and width of 
the enhancement are determined to be $M = 1812^{+19}_{-26}$ (stat) $\pm$ 
18 (syst) MeV/$c^2$ and $\Gamma = 105 \pm 20$ (stat) $\pm$ 28 (syst) MeV/$c^2$, 
and the product branching fraction is $\mathcal{B}(J/\psi\to\gamma X)\cdot
\mathcal{B}(X\to\omega\phi)$ = (2.61 $\pm$ 0.27 (stat) $\pm$  0.65 (syst)) 
$\times$ $10^{-4}$.  The mass and width of this state are not  compatible with 
any known scalars listed in the  Particle Data Group (PDG)~\cite{pdg04}. It 
could be an unconventional state \cite{liba,lixq,pedro,chaokt,bugg}. 
However, more statistics and  further 
studies are needed to clarify this.

\vspace {1cm}

The BES collaboration thanks the staff of BEPC and computing center for their 
hard efforts. This work is supported in part by the National Natural Science 
Foundation of China under contracts Nos. 10491300, 10225524, 10225525, 10425523, 
10521003, the Chinese Academy of Sciences under contract No. KJ 95T-03, the 
100 Talents Program of CAS under Contract Nos. U-11, U-24, U-25, and the 
Knowledge Innovation Project of CAS under Contract Nos. KJCX2-SW-N10, U-602, 
U-34 (IHEP), the National Natural Science Foundation of China under Contract 
No. 10225522 (Tsinghua University), and the Department of Energy under 
Contract No.DE-FG02-04ER41291 (U Hawaii).

\end{document}